\documentstyle[12pt,epsfig,amssymb]{article}
\begin{document}
\setlength{\baselineskip}{0.75cm}
\setlength{\parskip}{0.45cm}
\begin{titlepage}
\begin{flushright}
DO-TH 97/07 \linebreak
April 1997 \linebreak
(revised version, June 1997) 
\end{flushright}
\vskip 0.8in
\begin{center}
{\Large\bf Dark matter constraints on the parameter space \\ 
  and particle spectra in the nonminimal\\
  SUSY standard model}

\vspace{1.2cm}
\large
A.\ Stephan\\
\vspace{0.5cm}
\normalsize
Universit\"{a}t Dortmund, Institut f\"{u}r Physik, \\
\vspace{0.1cm}
D-44221 Dortmund, Germany \\
\vspace{1.6cm}
{\bf Abstract} 
\vspace{-0.3cm}
\end{center}

\noindent
We investigate the dark matter constraints for the 
nonminimal SUSY standard model (NMSSM).
The cosmologically restricted mass spectra of the NMSSM
are compared to the minimal SUSY standard model (MSSM).
The differences of the two models concerning the 
neutralino, sfermion and Higgs sector are discussed.
The dark matter condition leads to cosmologically allowed 
mass ranges for the SUSY particles in the NMSSM: 
$m_{\tilde{\chi}^0_1} \: < \: 300 \: GeV$,
$m_{\tilde{e}_R} \: < \: 300 \: GeV$,
$300 \: GeV \: < \: m_{\tilde{u}_R} \: < \: 1900 \: GeV$,
$200 \: GeV \: < \: m_{\tilde{t}_1} \: < \: 1500 \: GeV$,
$350 \: GeV \: < \: m_{\tilde{g}} \: < \: 2100 \: GeV$
and for the mass of the lightest scalar Higgs 
$m_{S_1} \: < \: 140 \: GeV$.
\end{titlepage}
%
%
\noindent
{\bf i) \it Introduction.}
\hspace*{.1cm}
Supersymmetry (SUSY) is suggested to solve the hierarchy problem
of the standard model (SM), if it is embedded 
in a grand unified theory (GUT).
The simplest SUSY extension of the SM is the 
minimal SUSY standard model (MSSM) \cite{nilles, haber}.
In the nonminimal SUSY standard model (NMSSM) \cite{ellis}
the Higgs sector of the MSSM is extended by a Higgs singlet N.
By introducing a Higgs singlet N the $\mu$-parameter of the
MSSM ($\mu$-problem) can be dynamically generated via
$\mu = \lambda x$ with the vaccum expectation value
(VEV) $\langle N \rangle = x$ and $\lambda$ being the 
Yukawa coupling of the Higgs fields.
The phenomenology of the NMSSM has been analysed in
several papers \cite{ellwglmin} - \cite{franke}.
SUSY models with Higgs singlets can be derived from superstring 
inspired $E_6$ or $SU(5) \times  U(1)$ GUT models
and they offer the possibility of spontaneous breaking 
of the CP symmetry.
The discrete ${\Bbb Z}_{\, 3}$ symmetry of the NMSSM causes a
domain wall problem, because the ${\Bbb Z}_{\, 3}$ symmetry
is spontaneously broken during the 
electoweak phase transition in the early universe. 
The domain wall problem of the NMSSM and possible solutions
to it are discussed in \cite{abelw}.
\newline
\noindent
In addition to the particle spectrum of the MSSM
there are two extra Higgses and one extra neutralino 
in the NMSSM, which can mix with the other Higgses 
and neutralinos, and thus modify their properties.
In many SUSY models the lightest SUSY particle (LSP)
is a neutralino, which is a good candidate for
cold dark matter.
In the NMSSM the LSP neutralino can have a larger 
singlino portion.
The cosmology of these LSP singlinos has been discussed 
in \cite{greene} at a given fixed low energy scale.
The experimentally and cosmologically allowed
parameter space of the NMSSM and the dark matter neutralinos
have been investigated in \cite{abeln} using RG evolutions.
\newline
\noindent
In the present paper we show, how the dark matter condition
restricts the mass spectra of the SUSY particles in the NMSSM.
The cosmologically restricted mass spectra of the MSSM and
NMSSM are compared and differences are discussed. 
In contrast to the previous investigation on 
NMSSM dark matter \cite{abeln},
we search for solutions which are the (global) minimum
of the effective 1 loop Higgs potential and
whose parameters are evolved with the SUSY RGEs
from the GUT scale to the electroweak scale
using the latest experimental constraints.
Furthermore we consider all possible decay channels
which occur in neutralino annihilation.

\noindent
{\bf ii) \it The NMSSM.}
\hspace*{.1cm} 
The NMSSM \cite{ellis} is a supersymmetric 
$SU(3)_C \times SU(2)_I \times U(1)_Y$ gauge theory
with two Higgs doublets $H_1$, $H_2$ and a Higgs singlet $N$.
This model is defined by the superpotential
(with only dimensionless couplings)
\begin{eqnarray}
   W = h_d \: H_1^T \: \epsilon \: \tilde{Q} \: \tilde{D} 
     - h_u \: H_2^T \: \epsilon \: \tilde{Q} \: \tilde{U} 
     + \lambda \: H_1^T \: \epsilon \: H_2 \: N 
     - \frac{1}{3} \: k \: N^3  
\end{eqnarray}
\noindent
($\epsilon$ is the antisymmetric tensor with $\epsilon_{1 2} \: = \: 1$)
and the SUSY soft breaking terms
\begin{eqnarray}
{\cal L}_{soft} & = &  - \, m^2_{H_1} \: |H_1|^2 - m^2_{H_2} \: |H_2|^2
   - m^2_N \: |N|^2 - M^2_Q \: |\tilde{Q}|^2
   - M^2_U \: |\tilde{U}|^2 - M^2_D \: |\tilde{D}|^2  \nonumber \\
   &  &  - \, h_d \: A_d \: H_1^T \: \epsilon \: \tilde{Q} \: \tilde{D} 
   + h_u \: A_u \: H_2^T \: \epsilon \: \tilde{Q} \: \tilde{U} 
   \nonumber \\
   &  &  + \, \lambda \: A_{\lambda} \: H_1^T \: \epsilon \: H_2 \: N 
   + \frac{1}{3} \: k \: A_k \: N^3  \nonumber \\ 
   &  &  + \, \frac{1}{2} \: M_3 \: \lambda_3 \: \lambda_3
   + \frac{1}{2} \: M_2 \: \lambda^a_2 \: \lambda^a_2
   + \frac{1}{2} \: M_1 \: \lambda_1 \: \lambda_1 + h.c. 
\end{eqnarray}
\noindent
The effective 1 loop Higgs potential consists of the
tree-level potential and the radiative corrections
\begin{eqnarray}
 V_{1 loop} = V_{tree} + V_{rad}. 
\end{eqnarray}
\noindent
The contributions of the top quark and stops $\tilde{t}_{1,2}$
are considered in the radiative corrections to the effective
Higgs potential \cite{ellwp}
\begin{eqnarray}
   V_{rad} = \frac{1}{64 \: \pi^2} \sum_{i} \: C_i \:
   (-1)^{2 S_i} \: (2 S_i + 1) \: m_i^4 \:
   \: \ln(\frac{m_i^2}{Q^2}), 
\end{eqnarray}
\noindent
where the sum is taken over all particles and antiparticles
with field-dependent mass $m_i$, spin $S_i$ and 
color degrees of freedom $C_i$.  
The electroweak gauge-symmetry $SU(2)_I \times U(1)_Y$
is spontaneously broken to the electromagnetic 
gauge-symmetry $U(1)_{em}$ by the Higgs VEVs
$\langle H_i^0 \rangle = v_i$  with $i = 1,2$ 
and $\langle N \rangle = x$.
The three minimum conditions for the VEVs have the form:
\begin{eqnarray}
   \frac{1}{2} \: m_Z^2 = 
   \frac{m_{H_1}^2 + \Sigma^1 
   - (m_{H_2}^2 + \Sigma^2) \: \tan^2 \beta }  
   {\tan^2 \beta - 1}
   - \lambda^2 x^2 
\end{eqnarray}
\begin{eqnarray}
   \sin(2 \beta) = 
   \frac{2 \: \lambda \: x \: (A_{\lambda} + k \: x)}
   {m_{H_1}^2 + \Sigma^1 + m_{H_2}^2 + \Sigma^2 
    + 2 \: \lambda^2 \: x^2 + \lambda^2 \: v^2} 
\end{eqnarray}
\begin{eqnarray}
   (m_N^2 + \Sigma^3) \: x^2 - k \: A_k \: x^3 + 2 \: k^2 \: x^4
   + \lambda^2 \: x^2 \: v^2
   - \frac{1}{2} \: (A_{\lambda} + 2 \: k \: x) \:
   \lambda \: x \: v^2 \: \sin(2 \beta) \: = \: 0
\end{eqnarray}
\noindent
where $v = \sqrt{v_1^2 + v_2^2} = 174 \: GeV$,  
$\tan \beta = v_2/v_1$ and 
$\Sigma^i = \partial V_{rad}/\partial v_i^2$.
\newline
\noindent
The mass of the neutralinos follows from the following part
of the lagrangian
\begin{eqnarray}
   {\cal L} = -\frac{1}{2} \Psi^T M \Psi + h.c.  
\end{eqnarray}
\begin{eqnarray}
   \Psi^T = (-i \lambda_1,-i \lambda^3_2,
   \Psi^0_{H_1},\Psi^0_{H_2},\Psi_N). 
\end{eqnarray}
\noindent
The symmetric mass matrix $M$ of the neutralinos in the basis 
given in (9) has the form:  
\[ \left( \begin{array}{ccccc}

   M_1  &  0  &  - m_Z \: \sin \theta_W \: \cos \beta  &
   m_Z \: \sin \theta_W \: \sin \beta  &  0  \\

   0  &  M_2  &  m_Z \: \cos \theta_W \: \cos \beta  &
   - m_Z \: \cos \theta_W \: \sin \beta  &  0  \\

   - m_Z \: \sin \theta_W \: \cos \beta  &  
   m_Z \: \cos \theta_W \: \cos \beta  &  0  &
   \lambda \: x  &  \lambda \: v_2  \\

   m_Z \: \sin \theta_W \: \sin \beta  & 
   - m_Z \: \cos \theta_W \: \sin \beta  &  
   \lambda \: x  &  0  &  \lambda \: v_1  \\
 
   0  &  0  &  \lambda \: v_2  &  \lambda \: v_1  &  - 2 \: k \: x  

   \end{array}\right). \]
\noindent
With $\mu = \lambda \: x$ the first $4 \times 4$ submatrix 
recovers the mass matrix of the MSSM \cite{haber}.  
The mass of the neutralinos is here obtained by diagonalising
the mass matrix $M$ with the orthogonal matrix $N$.
(Then some mass eigenvalues may be negative \cite{bartln}.)
\begin{eqnarray}
   {\cal L} = -\frac{1}{2} \: m_i \:
   \overline{\tilde{\chi}^0_i} \: \tilde{\chi}^0_i 
\end{eqnarray}
\begin{eqnarray}
   \tilde{\chi}^0_i = \left(\begin{array}{cc}
   \chi^0_i  \\ \overline{\chi}^0_i 
   \end{array}\right) \:\: \mbox{with} \:\:
   \chi^0_i = N_{ij} \Psi_j \:\:\: and \:\:\:
   M_{diag} = N \: M \: N^{T}. 
\end{eqnarray}
\noindent
Gauge coupling unification in non-supersymmetric GUTs
is already excluded by the precision measurements
at LEP \cite{ellisgut}.
In supersymmetric GUTs gauge coupling unification 
$g_a(M_X) = g_5 \simeq 0.72$ is possible 
at a scale $M_X \simeq 1.6 \times 10^{16} \: GeV$.
For the SUSY soft breaking parameters we suppose
universality at the GUT scale $M_X$:
\begin{eqnarray}
   \begin{array}{l}
 m_i(M_X) = m_0  \\
 M_a(M_X) = m_{1/2}  \\
 A_i(M_X) = A_0  \hspace*{1.5cm} (A_{\lambda}(M_X) = - A_0).   
   \end{array}
\end{eqnarray}
\noindent
The Yukawa couplings at the GUT scale take the values 
$\lambda(M_X) = \lambda_0$, $k(M_X) = k_0$ and $h_{t}(M_X) = h_{t 0}$.
With the SUSY RGEs \cite{savoy} the SUSY soft breaking parameters 
and the Yukawa couplings are evolved from the GUT scale 
to the electroweak scale $M_{weak} \simeq  100 \: GeV$. 
At $M_{weak}$ we minimize the effective 1 loop Higgs potential.
The 8 parameters of the NMSSM are $m_0$, $m_{1/2}$, $A_0$,
$\lambda_0$, $k_0$, $h_{t0}$, $\tan \beta$ and $x$,
with only 5 of them being independent,
which in our procedure are taken as  
$\lambda_0$, $k_0$, $h_{t0}$, $\tan \beta$ and $x$.
The remaining 3 parameters ($m_0$, $m_{1/2}$, $A_0$)
are numerically calculated from the three minimum conditions (5-7) 
following the method described in \cite{ellwtnb}:
\newline
Since we take here $\tan \beta$ and $x$ as (randomly chosen)
fixed input parameters, the three minimum conditions (5-7) depend 
only on the low energy parameters 
$m_i$, $A_i$, $\lambda$, $k$ and $h_t$    
generalized as $p_k(M_{weak})$: 
\begin{eqnarray}
   \frac{\partial \: V_{1 loop}}{\partial \: v_i} =
   F_i(p_k) = 0
   \: \: \: \: \: \: \: (i = 1-3).
\end{eqnarray}
The low energy parameters $\lambda$, $k$ and $h_{t}$
are fixed by the SUSY RGEs and the remaining (randomly chosen) 
fixed input parameters $\lambda_0$, $k_0$ and $h_{t 0}$.
With the SUSY RGEs the low energy parameters 
$m_i$ and $A_i$ are then only depending on the GUT scale parameters
$m_0$, $m_{1/2}$ and $A_0$.
Therefore all low energy parameters $p_k(M_{weak})$ can be written
as functions $f_k$ (which can be constant) of the GUT scale parameters
$m_0$, $m_{1/2}$ and $A_0$:
\begin{eqnarray}
   p_k(M_{weak}) = 
   f_k(m_0, m_{1/2}, A_0).
\end{eqnarray}
The functions $f_k$ are determined by numerically solving
the SUSY RGEs \cite{savoy}. 
In the three minimum conditions (13) we then substitute the 
low energy parameters $p_k(M_{weak})$ by the functions $f_k$
and obtain the equations
\begin{eqnarray}
   F_i(f_k(m_0, m_{1/2}, A_0)) = G_i(m_0, m_{1/2}, A_0) = 0
   \: \: \: \: \: \: \: (i = 1-3).
\end{eqnarray}
The SUSY breaking parameters $m_0$, $m_{1/2}$ and $A_0$
are then determined by numerically solving the equations
$G_i(m_0, m_{1/2}, A_0) = 0 \: \: (i = 1-3)$.
\newline
Because we assume $h_t \gg h_b, h_{\tau}$,
we restrict $|\tan \beta \: |$ to be smaller than 20 \cite{ellwtnb}.
For the singlet VEV $x$ we choose the range
$|x| < 80 \: TeV$ \cite{ellwtnb, ellwn, king}. 
\newline
\noindent
The physical minimum of the Higgs potential
should be the global minimum.
We check, whether the physical minimum
is lower than the unphysical minima of 
$V_{1 loop}(v_1, v_2, x)$ with at least one 
vanishing VEV \cite{ellwglmin}.
\newline
\noindent
The conditions \cite{savoy}, that the minimum of the scalar potential
does not break the conservation of charge and colour,
can be formulated as follows:
\begin{eqnarray}
  A_{u_i}^2 \leq 3 (m_{H_2}^2 + m_{Q_i}^2 + m_{U_i}^2 ) \: \:
  \mbox{at scale} \: \: Q \sim A_{u_i}/h_{u_i}          \\    
  A_{d_i}^2 \leq 3 (m_{H_1}^2 + m_{Q_i}^2 + m_{D_i}^2 ) \: \:      
  \mbox{at scale} \: \: Q \sim A_{d_i}/h_{d_i}          \\
  A_{e_i}^2 \leq 3 (m_{H_1}^2 + m_{L_i}^2 + m_{E_i}^2 ) \: \:     
  \mbox{at scale} \: \: Q \sim A_{e_i}/h_{e_i}.
\end{eqnarray}
\noindent
For the top quark pole mass we take the range 
$169 \: GeV < m_t < 181 \: GeV$ corresponding to the 
recent world average measurement \cite{top}. 
The SUSY particles in the NMSSM have to fulfill the following
experimental conditions:
$ m_{\tilde{\nu}} \geq 41.8 \: GeV \:\: \cite{PD} $,
$ m_{\tilde{e}} \geq 65 \: GeV \:\: \cite{pALEPH} $,
$ m_{\tilde{q}} \geq 176 \: GeV 
   \:\: ( \mbox{if} \: \: m_{\tilde{g}} < 300 \: GeV ) 
   \:\: \cite{PD} $ and
$ m_{\tilde{q}} \geq 70 \: GeV \:\: ( \mbox{for all} 
  \: \: m_{\tilde{g}} ) \:\: \cite{bisset} $,
\newline
$ m_{\tilde{t}} \geq 58.3 \: GeV \:\: \cite{pL3} $,
$ m_{\tilde{g}} \geq 173 \: GeV \:\: \cite{FERMILAB} $,
$ m_{\tilde{\chi}^+} \geq 71.3 \: GeV \:\: \cite{pDELPHI} $,
$ m_{H^+} \geq 43.5 \: GeV \:\: \cite{PD} $,
\newline
$m_{S_1} \geq 70.7 \: GeV$ \cite{pALEPH} or the coupling 
of the lightest scalar Higgs $S_1$ to the Z boson 
is reduced compared to the SM \cite{pALEPH, L3}
assuming visible decays with branching ratios
like the SM Higgs \cite{ellwn}, and
\[ \sum_{i,j} \Gamma(Z \rightarrow 
   \tilde{\chi}^0_i \tilde{\chi}^0_j) < 30 \: MeV \:\:
   \cite{diaz} \]
\[ \Gamma(Z \rightarrow 
   \tilde{\chi}^0_1 \tilde{\chi}^0_1) < 7 \: MeV \:\:
   \cite{diaz} \]
\[ BR(Z \rightarrow 
   \tilde{\chi}^0_i \tilde{\chi}^0_j) < 10^{-5} 
   \:\:,\: (i,j) \not= (1,1) \:\: \cite{diaz}. \]

\noindent
{\bf iii) \it Dark matter constraints.}
\hspace*{.1cm}
Inflationary cosmological models suggest a flat universe
with $\Omega = \rho / \rho_{crit} = 1$.
Big bang nucleosynthesis restricts the 
baryon density to $\Omega_{baryons} \leq 0.1$.
The missing matter is called dark matter.
One of the favoured theories to explain the structure formation
of the universe is the cold + hot dark matter model (CHDM)
\cite{davis}.
In this model the dark matter consists of 
hot dark matter (massiv neutrinos)
and cold dark matter (neutralinos) with
$\Omega_{hot} \simeq 0.3$ and $\Omega_{cold} \simeq 0.65$.
With the Hubble constant $H_0 = 100 \: h_0 \: km \: s^{-1} \: Mpc^{-1}$
and the range $0.4 < h_0 < 1$, the condition
for the cosmic density of the neutralinos in the CHDM model reads
\begin{eqnarray}
   0.1 \leq \Omega_{\chi} h_0^2 \leq 0.65. 
\end{eqnarray}
\noindent
With the annihilation cross section of the neutralinos
$ \sigma_{ann} v = a + b \: v^2 $ 
\hspace*{.07cm} ($v$ is the relative velocity 
of the neutralinos) 
the cosmic neutralino density \cite{drees, kolb, griest}
can be calculated
\begin{eqnarray}
   \Omega_{\chi} h_0^2 = \frac{1.07 \times 10^9 \: x_{fr}}
   {\sqrt{g_{\ast}} \: M_P \: (a + \frac{3 \: b}{x_{fr}})} 
   \frac{1}{GeV}, 
\end{eqnarray}
\noindent
where $M_P = 1.221 \times 10^{19} \: GeV$ is the Planck mass
and $g_{\ast} \approx 81$ is the effective number of degrees
of freedom at $T_{fr}$.
The freeze-out temperature of the neutralinos $T_{fr}$
follows from the equation \cite{drees, kolb, griest}
\begin{eqnarray}
   x_{fr} = \ln \left(0.0764 \: |m_{\tilde{\chi}^0_1}| \: M_P \:
   \frac{a + \frac{6 \: b}{x_{fr}}}{\sqrt{g_{\ast} \: x_{fr}}} 
   \:c\:(2 + c)\right)  
\end{eqnarray}
\noindent
with $x_{fr} = |m_{\tilde{\chi}^0_1}|/T_{fr}$ and $c = 1/2$.
\newline
\noindent
In the MSSM the formulas for the annihilation cross section
of the neutralinos into the different decay products $X$ and $Y$
can be found in \cite{drees}
\begin{eqnarray}
   \sigma_{ann}(\tilde{\chi}^0_1 \: \tilde{\chi}^0_1 
   \rightarrow X \: Y) v = \frac{1}{4} 
   \frac{\overline{\beta}_f}{8 \pi s S}
   \left[|A(^1S_0)|^2 + \frac{1}{3} 
   (|A(^3P_0)|^2 + |A(^3P_1)|^2 + |A(^3P_2)|^2 ) \right] 
\end{eqnarray}
\begin{eqnarray}
   \overline{\beta}_f = \sqrt{ 1 
   - \frac{2 (m_X^2 + m_Y^2)}{s} + \frac{(m_X^2 - m_Y^2)^2}{s^2} } 
\end{eqnarray}
\noindent
$s \approx 4 m_{\tilde{\chi}}^2$ is the center of mass energy squared.
S is a symmetry factor, which is 2 if $X = Y$. 
The partial wave amplitude $A$ describes annihilation
from an initial state $^{2 S + 1}L_J$ with spin S,
orbital angular momentum L and total angular momentum J.
\newline
\noindent
In the NMSSM the annihilation cross section of the neutralinos
changes because of more particles in this model, which can mix, 
and modified vertices. 
In our numerical analysis we consider all possible decay channels.
The relevant formulas for the NMSSM are obtained by substituting
the MSSM couplings in \cite{drees} by the corresponding NMSSM couplings 
and by slightly modifying the partial wave amplitudes in \cite{drees} 
and will be given in \cite{ich}.
Here we only give the NMSSM couplings for  
the dominant decay channel  
\[ \tilde{\chi}^0_1 \: \: \tilde{\chi}^0_1 \: \rightarrow \:
   f_a \: \: \overline{f_a}. \]
\noindent
The fermions are produced by Z boson, scalar Higgs $S_i \: (i = 1-3)$
and pseudoscalar Higgs $P_{\alpha} \: (\alpha = 1, 2)$ exchange in the 
s-channel and sfermion $\tilde{f}_{1, 2}$ exchange in the t- and u-channel.
When the LSP neutralino, which is denoted with index 0 in \cite{drees} 
instead of 1, is predominantly a bino (${N_{0 1}}^2 \: \approx \: 1$
in eq. (11)), the annihilation cross section is dominated 
by sfermion exchange. 
The couplings of the unmixed sfermion $\tilde{f}_{L, R}$ 
to the fermion $f_a$ and LSP neutralino, are given by
\begin{eqnarray}
   X_{a 0} = - \sqrt{2} \, g_2 \, [T_{3 a} \, N_{0 2}
   - \tan \theta_W \, (T_{3 a} - e_{f_a}) \, N_{0 1}] 
\end{eqnarray}
\begin{eqnarray}
   Y_{a 0} = \sqrt{2} \, g_2 \, \tan \theta_W \, e_{f_a} \, N_{0 1} 
\end{eqnarray}
\begin{eqnarray}
   Z_{u 0} = - \frac{g_2 \, m_u}{\sqrt{2} \, \sin \beta \, m_W } \, 
   N_{0 4} 
\end{eqnarray}
\begin{eqnarray}
   Z_{d 0} = - \frac{g_2 \, m_d}{\sqrt{2} \, \cos \beta \, m_W } \, 
   N_{0 3} 
\end{eqnarray}
and refer to eqs. (A32a)-(A32c) in \cite{drees}. 
These couplings have the same form in the MSSM and NMSSM.
The only difference follows from the matrix $N$,
which diagonalizes the mass matrix of the neutralinos.
The couplings $X'_{a0}$, $W'_{a0}$, $Z'_{a0}$ and $Y'_{a0}$
of the mixed sfermion $\tilde{f}_{1, 2}$ to the fermion $f_a$ 
and LSP neutralino are easily given in terms of the unmixed case
as described by eq. (A31) in \cite{drees}.
\noindent
When the Higgsino portion of the LSP neutralino is larger
(larger ${N_{0 3}}^2$, ${N_{0 4}}^2$ in eq. (11)),
Z boson exchange becomes important for the annihilation cross section. 
The coupling of the Z boson to the LSP neutralinos $O''^L_{0 0}$ 
(eq. (A5e) in \cite{drees}) has the same form in the MSSM 
and NMSSM, but a different matrix $N$ enters: 
\begin{eqnarray}
   O''^L_{0 0} = - \frac{1}{2} \, {N_{0 3}}^2
   + \frac{1}{2} \, {N_{0 4}}^2. 
\end{eqnarray}
In the NMSSM the Higgs-couplings to the fermions have to be
modified compared to the MSSM, 
\begin{eqnarray}
  h_{P_{\alpha} u} = 
  - \frac{g_2 \: m_u \: U_{{\alpha} 2}^P}{2 \: \sin \beta \: m_w} 
  \hspace*{2.cm}
  h_{P_{\alpha} d} = 
  - \frac{g_2 \: m_d \: U_{{\alpha} 1}^P}{2 \: \cos \beta \: m_w} 
\end{eqnarray}
\begin{eqnarray}
  h_{S_i u} = - \frac{g_2 \: m_u \: U_{i 2}^S}{2 \: \sin \beta \: m_w}  
  \hspace*{2.cm}
  h_{S_i d} = - \frac{g_2 \: m_d \: U_{i 1}^S}{2 \: \cos \beta \: m_w}. 
\end{eqnarray}
\noindent
The $3 \times 3$ mass matrix of the scalar Higgses $S_i$ \cite{ellis} 
in the basis ($H^0_{1 R}$, $H^0_{2 R}$, $N_R$) with 
$H^0_{i R} \: = \: \sqrt{2} \: Re\{H^0_i\}$ (i = 1-3)
is diagonalised by the matrix $U^S$.
The $3 \times 3$ mass matrix of the pseudoscalar Higgses $P_{\alpha}$ 
and the Goldstone boson \cite{ellis} in the basis 
($H^0_{1 I}$, $H^0_{2 I}$, $N_I$) with 
$H^0_{i I} \: = \: \sqrt{2} \: Im\{H^0_i\}$ (i = 1-3)
is here diagonalised by the $3 \times 3$ matrix $U^P$.
These NMSSM couplings have to be substituted for the MSSM couplings
given in eqs. (A33a)-(A33c) of \cite{drees}.
The coupling of the scalar Higgs $S_a$ to the neutralinos 
is different in the NMSSM and the MSSM.
In the NMSSM this coupling can be written as
\begin{eqnarray}
   T_{S_a i j} = - U_{a 1}^S \: Q''_{i j} + U_{a 2}^S \: S''_{i j}
   + U_{a 3}^S \: Z''_{i j} 
\end{eqnarray}
\begin{eqnarray}
   Q''_{i j} = \frac{1}{2 \: g_2} [ N_{i 3} \: 
   (g_2 \: N_{j 2} - g_y \: N_{j 1})
   + \sqrt{2} \: \lambda \: N_{i 4} \: N_{j 5} + (i \leftrightarrow j) ] 
\end{eqnarray}
\begin{eqnarray}
   S''_{i j} = \frac{1}{2 \: g_2} [ N_{i 4} \: 
   (g_2 \: N_{j 2} - g_y \: N_{j 1})
   - \sqrt{2} \: \lambda \: N_{i 3} \: N_{j 5} + (i \leftrightarrow j) ] 
\end{eqnarray}
\begin{eqnarray}
   Z''_{i j} = \frac{1}{2 \: g_2} 
   [ - \sqrt{2} \: \lambda \: N_{i 3} \: N_{j 4} 
   + \sqrt{2} \: k \: N_{i 5} \: N_{j 5} + (i \leftrightarrow j) ], 
\end{eqnarray}
\noindent
to be substituted for the MSSM couplings given in eqs. (A9b,A9c)
of \cite{drees}. 
The coupling of the pseudoscalar Higgs $P_{\alpha}$ to the neutralinos 
also changes.
In the NMSSM this coupling has the following form 
\begin{eqnarray}
   T_{P_{\alpha} i j} = - U_{\alpha 1}^P \: Q'''_{i j} 
   + U_{\alpha 2}^P \: S'''_{i j} - U_{\alpha 3}^P \: Z'''_{i j} 
\end{eqnarray}
\begin{eqnarray}
   Q'''_{i j} = \frac{1}{2 \: g_2}  
   [ N_{i 3} \: (g_2 \: N_{j 2} - g_y \: N_{j 1})
   - \sqrt{2} \: \lambda \: N_{i 4} \: N_{j 5} + (i \leftrightarrow j) ] 
\end{eqnarray}
\begin{eqnarray}
   S'''_{i j} = \frac{1}{2 \: g_2} 
   [ N_{i 4} \: (g_2 \: N_{j 2} - g_y \: N_{j 1})
   + \sqrt{2} \: \lambda \: N_{i 3} \: N_{j 5} + (i \leftrightarrow j) ] 
\end{eqnarray}
\begin{eqnarray}
   Z'''_{i j} = Z''_{i j},  
\end{eqnarray}
\noindent
which has to be used instead of the MSSM coupling in eq. (A9a)
of \cite{drees}.

\noindent
{\bf iv) \it Particle spectra.}
\hspace*{.1cm}
To obtain the mass spectra of the SUSY particles in the NMSSM,
we randomly generate $\sim 5.5 \times 10^8$ points in the
5 dimensional parameter space of the input parameters
$\lambda_0$, $k_0$, $h_{t0}$, $\tan \beta$ and $x$.
The SUSY breaking parameters $m_0$, $m_{1/2}$ and $A_0$ 
are then determined by numerically solving the three 
minimum equations (15) as described there.  
Having found a solution to the SUSY RGEs,
whose effective 1 loop Higgs potential has a local minimum
for the input parameters $\tan \beta$ and $x$, we calculate the masses 
of all SUSY particles \cite{haber, bartl} and Higgses 
\cite{ellis, ellwp, elli2, king} and impose the additional theoretical 
and experimental constraints described above in ii).
With all these requirements, about $4900$ solutions remain,
from which about $2000$ solutions are cosmologically acceptable.
\noindent
For the constrained MSSM in \cite{kolda} a scatter-plot of 
solutions in the $(m_{1/2}, m_0)$ plane is shown.
Upper bounds of $1.1 \: TeV$ for $m_{1/2}$ 
and $600 \: GeV$ for $m_0$ are found.
For $m_{1/2} \approx 100 \: GeV$ solutions with 
much larger $m_0$ are allowed due to
Z-pole enhanced neutralino pair annihilation 
($m_{\tilde{\chi}^0_1} \approx m_Z/2$).
The comparable scatter-plot for the NMSSM 
(more details in \cite{ich})
shows only a few cosmologically allowed solutions 
with $m_0 \gg m_{1/2}$.
The reason, why the case $m_0 \gg m_{1/2}$ is somewhat 
suppressed in the NMSSM, is discussed in \cite{ellws}.
It follows from the condition, that the minimum of the 
scalar potential does not break the conservation 
of charge and colour,  
and from the condition that the physical minimum
is lower than the symmetric vacuum, $A_0^2 > 9 m_0^2$.

\begin{figure}
\vspace*{-2.cm}
\hspace*{-.5cm}
\epsfig{file=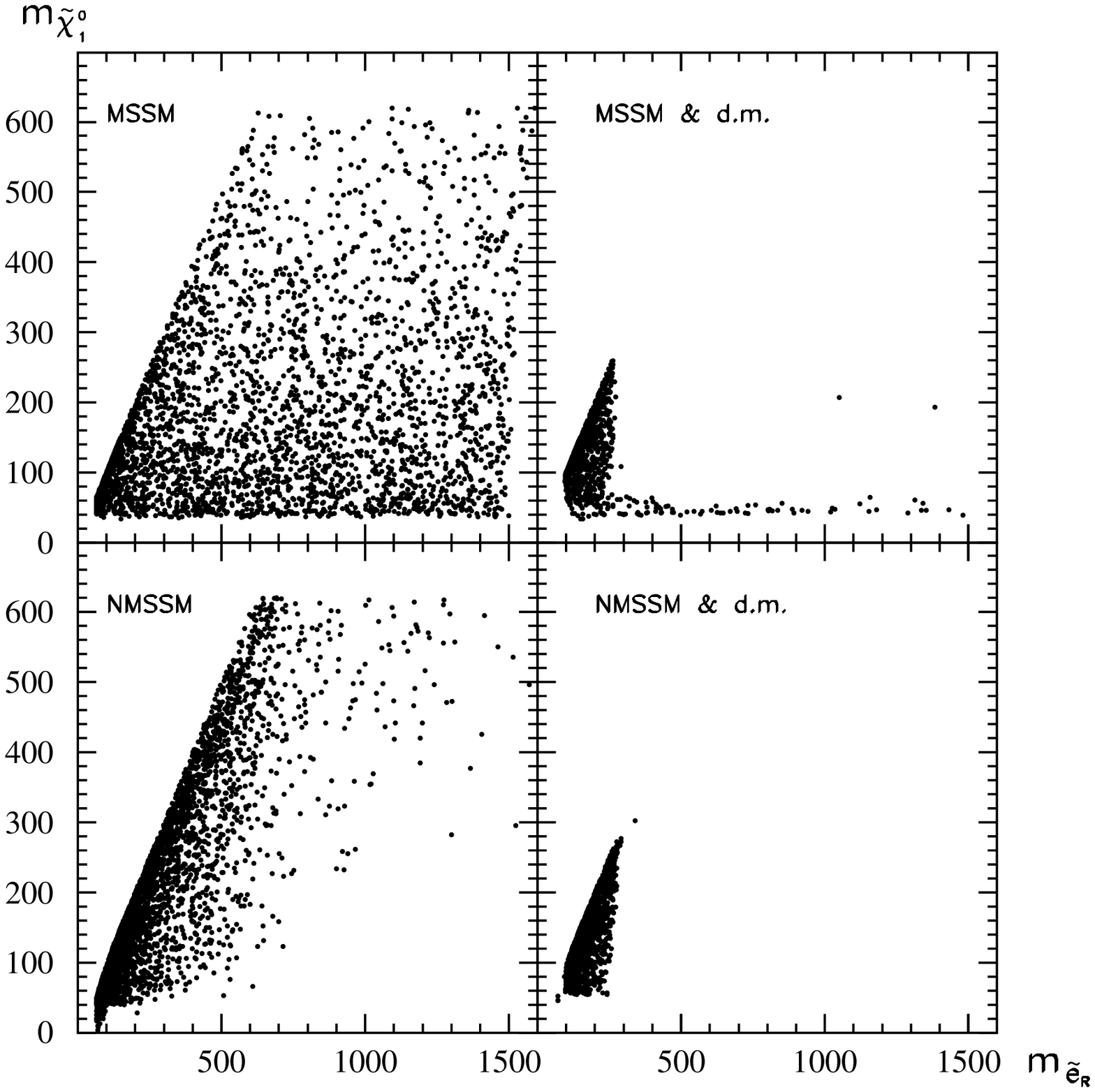,width=17cm}
\vspace*{1.cm}
\begin{center}
\refstepcounter{figure}
\label{ne}
{\normalsize\bf Fig.\ \thefigure}
\end{center}
\end{figure}

\noindent
As a few representative examples for the importance
of the dark matter constraints we show in the following 
some scatter-plots for the mass of SUSY particles.
In the upper left picture of fig.\ref{ne} 
the mass of the LSP neutralino versus the mass of
the lighter selectron is shown for the MSSM.
In the MSSM the LSP neutralino is mostly a bino.
In the upper right picture the dark matter condition 
is imposed and gives un upper bound of $280 \: GeV$
for the LSP neutralino and a cosmologically allowed 
mass range of $100 \: GeV \: < \: m_{\tilde{e}_R} \: < \: 300 \: GeV$
for the lighter selectron.
These bounds for the LSP neutralino and lighter selectron
can be understood from the fact \cite{drees}, that for a 
bino-like LSP neutralino the annihilation cross section
is dominated by sfermion exchange. 
Heavier selectrons are cosmologically allowed,
if Z-pole or Higgs-pole enhanced neutralino 
pair annihilation \cite{kolda} occurs.

\noindent
For the NMSSM the lower left picture of fig.\ref{ne}
shows the mass of the LSP neutralino versus the mass of
the lighter selectron.
Compared to the MSSM there are not so many solutions
in the NMSSM with small LSP neutralino mass 
and large selectron mass,
which follows from the suppression of
$m_0 \gg m_{1/2}$ in the NMSSM \cite{ellws}. 
In the NMSSM some LSP neutralinos can have a larger singlino portion.
Most of these singlinos are decoupled pure singlinos 
\cite{ellws, king}, but a smaller fraction of the singlinos 
can be mixed states (\cite{greene}, more details in \cite{ich}).
The LSP neutralinos with mass below $40 \: GeV$ and 
$m_{\tilde{e}_R} \: \simeq \: 70 \: GeV$ in the lower left picture 
of fig.\ref{ne} are lighter singlinos.
\newline
\noindent
The lower right picture of fig.\ref{ne} shows the
NMSSM solutions, which survive the dark matter condition.
The cosmologically allowed solutions in the MSSM and NMSSM
look quite similar except that heavier selectrons are 
allowed in the former. 
With the dark matter condition nearly all decoupled pure
LSP singlinos can be excluded (more details in \cite{ich}),
what can be seen for the lighter singlinos 
($m_{\tilde{\chi}^0_1} \: < \: 40 \: GeV \: \: \: \: \mbox{with} 
\: \: \: \: m_{\tilde{e}_R} \: \simeq \: 70 \: GeV$)
by comparing the lower pictures of fig.\ref{ne}.
\newline
Another difference between the MSSM and NMSSM is the enhanced 
lower bound of $50 \: GeV$ for most of the LSP neutralinos in the NMSSM.
The dark matter condition in the NMSSM in fig.\ref{ne} exludes nearly 
all LSP neutralinos with mass $m_{\tilde{\chi}^0_1} \: \approx 
\: m_Z/2$, even if they are not singlinos. 
The reason for this difference is the larger coupling $O''^L_{0 0}$
in eq. (28) of the Z boson to the LSP neutralinos in the NMSSM
corresponding to neutralinos with larger Higgsino portion.
In the MSSM the coupling $O''^L_{0 0}$ can be very small for
nearly pure bino-like LSP neutralinos, so that the neutralino 
pair annihilation becomes not too large from Z exchange
near the Z-pole.
The difference follows from the circumstance,
that the NMSSM can be obtained from the MSSM in the limit
$\lambda, k \: \rightarrow 0$, $| \, x \, | \rightarrow \infty$
with $\lambda \: x$ and $k \: x$ fixed.
In this limit the first two minimum conditions (5),(6) of the NMSSM
correspond to the two minimum conditions of the MSSM,
whereas the third minimum condition (7) of the NMSSM 
would be an extra constraint for the MSSM \cite{ellwtnb,king}.
The third minimum condition restricts the parameter region
of the NMSSM compared to the MSSM in such a way, that the 
coupling $O''^L_{0 0}$ of the Z boson to the LSP neutralinos  
is larger in the NMSSM for $m_{\tilde{\chi}^0_1} \: \approx \: m_Z/2$.

\begin{figure}
\vspace*{-3.cm}
\hspace*{-.5cm}
\epsfig{file=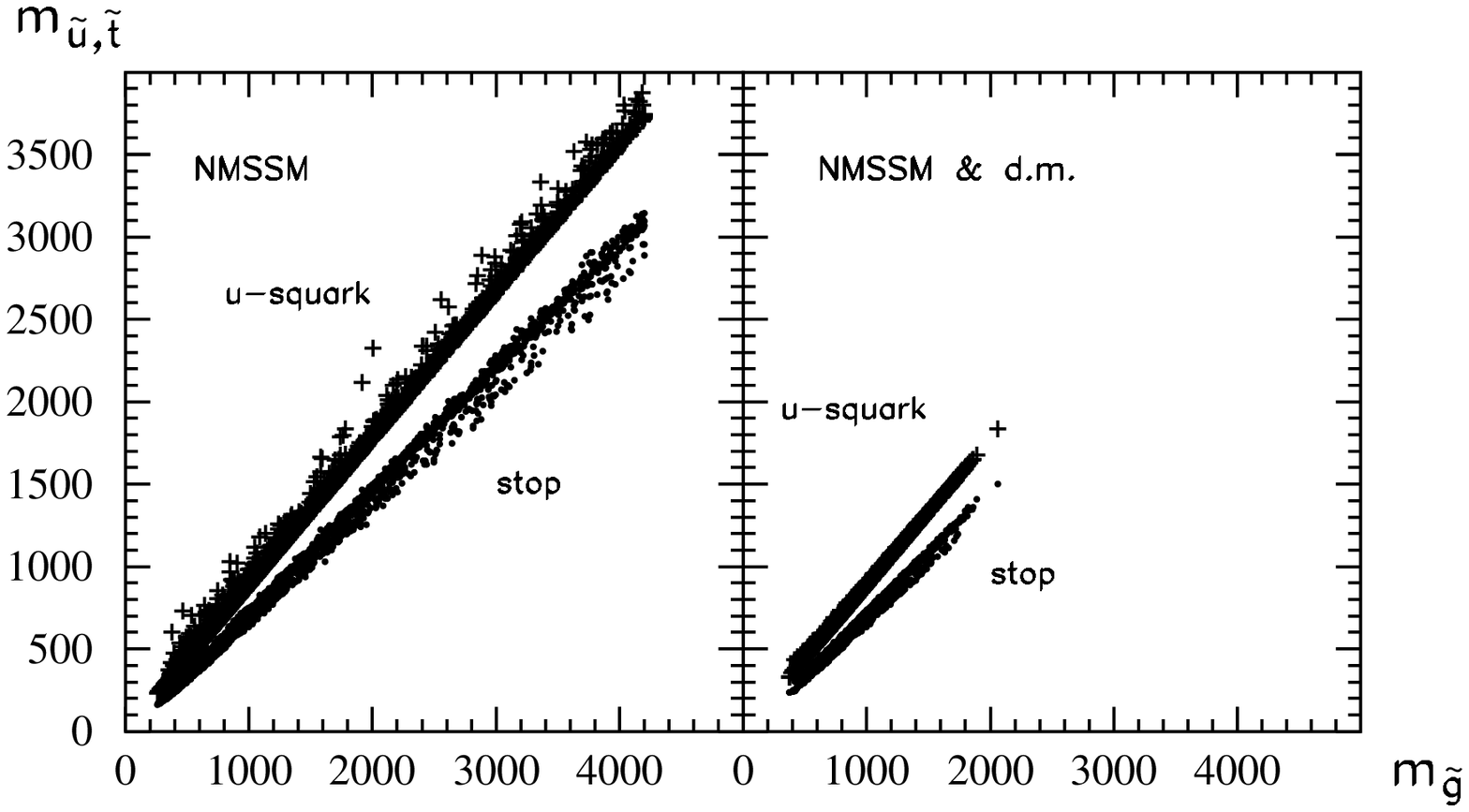,width=17cm}
\vspace*{-2.2cm}
\begin{center}
\refstepcounter{figure}
\label{utg}
{\normalsize\bf Fig.\ \thefigure}
\end{center}
\end{figure}

\noindent
In the left picture of fig.\ref{utg} we show the mass of the
lighter u-squark and the lighter stop versus the mass of the gluino 
for the NMSSM. 
In contrast to the MSSM \cite{barger,ich}, 
the squark masses in the NMSSM
are nearly proportional to the gluino mass.
The already mentioned reason is, that in the NMSSM 
solutions with $m_0 \gg m_{1/2}$ 
are somewhat disfavoured.
In the right picture of fig.\ref{utg}
the cosmologically allowed lighter u-squark and 
lighter stop masses are shown.
In contrast to the MSSM, the dark matter condition improves
the lower bound for the gluino to $350 \: GeV$ in the NMSSM.
The reason is the enhanced lower bound of $50 \: GeV$ for 
most of the LSP neutralinos in fig.\ref{ne}.
Furthermore, the Higgsino or singlino portion of the lightest 
cosmologically allowed LSP neutralinos is larger in the NMSSM 
than in the MSSM as already discussed.
For these LSP neutralinos with larger Higgsino or singlino portion, 
the gaugino mass $M_1$ is larger than their mass $m_{\tilde{\chi}^0_1}$.
A larger gaugino mass $M_1$ leads to a larger gluino mass $M_3$
in the NMSSM.
Because the mass of the squarks are nearly proportional to 
the gluino mass in the NMSSM, the lower bound for the 
lighter u-squark improves to $300 \: GeV$ and the lower bound 
for the lighter stop improves to $200 \: GeV$.
This is in contrast to the MSSM, where the dark matter condition
does not change the lower bounds. 
The cosmologically upper bound of $1900 \: GeV$ for the 
lighter u-squark, $1500 \: GeV$ for the lighter stop and 
$2100 \: GeV$ for the gluino in the NMSSM
are similar to the corresponding upper bounds in the MSSM 
and are connected with the cosmologically upper bound 
of $300 \: GeV$ for the lighter selectron and the 
bino-like LSP neutralino.

\begin{figure}
\vspace*{-3.cm}
\hspace*{-.5cm}
\epsfig{file=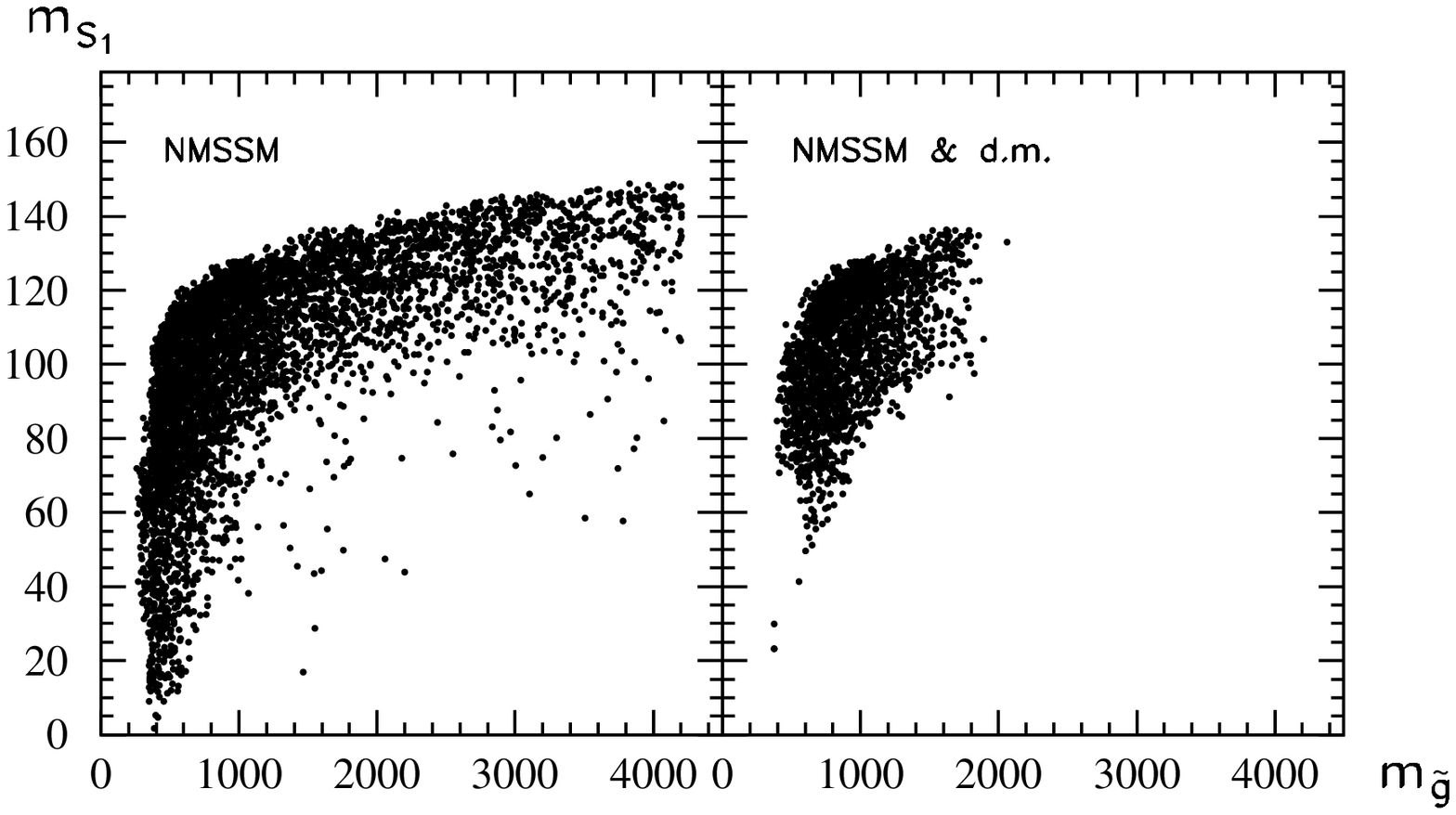,width=17cm}
\vspace*{-2.2cm}
\begin{center}
\refstepcounter{figure}
\label{sg}
{\normalsize\bf Fig.\ \thefigure}
\end{center}
\end{figure}

\noindent
In the left picture of fig.\ref{sg} the mass of the
lightest scalar Higgs $S_1$ versus the mass 
of the gluino is shown for the NMSSM. 
The lightest scalar Higgs can be lighter than 
the MSSM Higgs bound of $62.5 \: GeV$ \cite{pALEPH}.
In this case the lightest scalar Higgs is predominantly
a Higgs singlet $N$ \cite{ellwn, elli1, elli2}.
The right picture of fig.\ref{sg} shows
only solutions, which fulfill the dark matter condition. 
For bino-like LSP neutralinos the cosmologically allowed 
mass range for the lightest scalar Higgs $S_1$ is
$40 \: GeV \: < \: m_{S_1} \: < \: 140 \: GeV$,
which follows from the cosmologically allowed mass range 
for the lighter selectron of
$100 \: GeV \: < \: m_{\tilde{e}_R} \: < \: 300 \: GeV$.
Very light scalar Higgses $S_1$ can be cosmologically allowed,
if the LSP neutralino is a mixed singlino \cite{greene} or 
a nearly pure singlino with mass 
$m_{\tilde{\chi}^0_1} \: = \: m_Z/2$,
so that Z-pole enhanced neutralino pair annihilation occurs.
The corresponding plots in the MSSM \cite{boer,ich} look very similar
with the exception, that there are no Higgses below
the bound of $62.5 \: GeV$ \cite{pALEPH}.

\noindent
{\bf v) \it Conclusions.}
\hspace*{.1cm}
We have investigated the particle spectra of the MSSM and NMSSM
without and with the dark matter condition.
In contrast to the MSSM, the squark masses in the NMSSM
are nearly proportional to the gluino mass.
We found, that pure LSP singlinos and most of the 
light scalar Higgs singlets are cosmologically excluded.
With the dark matter condition the following cosmologically allowed 
mass ranges for the SUSY particles in the NMSSM can be given:
$m_{\tilde{\chi}^0_1} \: < \: 300 \: GeV$,
$m_{\tilde{e}_R} \: < \: 300 \: GeV$,
$300 \: GeV \: < \: m_{\tilde{u}_R} \: < \: 1900 \: GeV$,
$200 \: GeV \: < \: m_{\tilde{t}_1} \: < \: 1500 \: GeV$,
$350 \: GeV \: < \: m_{\tilde{g}} \: < \: 2100 \: GeV$,
$m_{S_1} \: < \: 140 \: GeV$.
The upper bounds for the mass of the SUSY particles in the NMSSM 
are comparable to the MSSM bounds.
The larger Higgsino or singlino portion of the lightest LSP neutralinos 
in the NMSSM leads to  improved lower bounds 
for the gluinos and squarks compared to the MSSM.

\newpage
\noindent
I am grateful to M. Gl\"uck and E. Reya 
for suggestions and many helpful discussions.
Also I would like to thank U. Ellwanger and C.A. Savoy
for useful conversations.
The work has been supported by the 
'Graduiertenkolleg am Institut f\"ur Physik
der Universit\"at Dortmund'.


\section*{Figure Captions}
\begin{description}

\item[Fig\ 1] The mass of the LSP neutralino versus
the lighter selectron mass. The upper pictures are for the MSSM,
the lower pictures are for the NMSSM. In the right pictures 
the dark matter condition is imposed.

\item[Fig\ 2] The mass of the lighter u-squark and 
the lighter stop versus the gluino mass in the NMSSM. 
In the right picture the dark matter condition is imposed.

\item[Fig\ 3] The mass of the lightest Higgs $S_1$ versus
the gluino mass in the NMSSM. In the right picture 
the dark matter condition is imposed.

\end{description}
\newpage

\newpage

\newpage

\end{document}